| | |
|---|---|
| **Title** | **Influence of ambient air on the flowing afterglow of an atmospheric pressure Ar/$O_2$ radiofrequency plasma** |
| **Authors** | C. Y. Duluard, T. Dufour, J. Hubert, and F. Reniers |
| **Affiliations** | Université Libre de Bruxelles, Faculté des Sciences, Chimie Analytique et Chimie des Interfaces, Boulevard du Triomphe, 2, 1050 Bruxelles, Belgium |
| **Ref.** |  |
| **DOI** |  |
| **Abstract** | The influence of ambient air on the flowing afterglow of an atmospheric pressure Ar/$O_2$ radiofrequency plasma has been investigated experimentally. Spatially resolved mass spectrometry and laser induced fluorescence on OH radicals were used to estimate the intrusion of air in between the plasma torch and the substrate as a function of the torch-to-substrate separation distance. No air is detected, within the limits of measurement uncertainties, for separation distances smaller than 5 mm. For larger distances, the effect of ambient air can no longer be neglected, and radial gradients in the concentrations of species appear. The Ar 4p population, determined through absolute optical emission spectroscopy, is seen to decrease with separation distance, whereas a rise in emission from the $N_2$(C–B) system is measured. The observed decay in Ar 4p and $N_2$(C) populations for separation distances greater than 9mm is partly assigned to the increasing collisional quenching rate by $N_2$ and $O_2$ molecules from the entrained air. Absorption measurements also point to the formation of ozone at concentrations from $10^{14}$ to $10^{15}$ $cm^{-3}$, depending both on the injected $O_2$ flow rate and the torch-to-substrate separation distance. |

# 1. Introduction

Non-equilibrium atmospheric pressure plasmas have recently created widespread interest (cf. dedicated topics in symposia on plasma and material science), notably for thin film deposition,1–3 surface treatment,4–6 and biomedical applications.7–9 Their attractiveness stems principally from both a non-reliance on elaborate vacuum technology and their low operating gas temperatures.

The atmospheric pressure plasma under study in this paper is that of an AtomfloTM 250D plasma torch from Surfx Technologies LLC, commercialized for plasma enhanced chemical vapor deposition1,3 and for treating polymer surfaces to increase, for instance, their adhesion properties.6 The plasma is a capacitive discharge excited by a 27.12MHz radiofrequency (RF) power source. Substrates to be treated are placed downstream of the plasma. This creates a remote plasma operating regime where reactions occur quasi-exclusively between neutral reactive species and the substrate surface. Helium is normally the main feedstock gas, but this has been replaced with argon in this study. With the ever increasing rarity and expected rises in the price of helium, the choice to use argon has been made to prepare for the possibility that helium becomes less commercially attractive.

For material processing applications, a crucial issue is the control of the reactive species impinging the substrate surface. In this context, the aim of this study is to quantify the effect of ambient air on the physical and chemical properties of the flowing afterglow. To that purpose, a characterization of the plasma afterglow as a function of the plasma torch-to-substrate separation distance is undertaken by optical diagnostics and mass spectrometry. Optical emission and absorption spectroscopy are employed to gain insight into the variations of the Ar 4p population and the creation of new species arising from the interaction between the plasma afterglow and the air. Both spatially resolved mass spectrometry and laser induced fluorescence (LIF) on OH radicals are implemented to determine the intrusion of air in between the plasma torch and the substrate. As a follow-up to the results obtained, we propose possible mechanisms for the changes in the properties of the afterglow induced by air entrainment. We also show that for short torch-to-substrate separation distances (<5 mm), the plasma afterglow is effectively free of any contamination from the ambient air.

# 2. Experimental setup & Diagnostics

## 2.1. Plasma source

Figure 1 presents a schematic view of the plasma torch under study. The discharge is produced between two perforated, parallel plate electrodes, 2.5 cm in diameter. The upper electrode is powered at a frequency of 27.12 MHz, while the lower electrode is grounded. The Ar and $O_2$ gases flow through the electrodes at rates of 30 L/min and in the range 0–30 mL/min, respectively. Additionally, precursors for plasma enhanced chemical vapor deposition can be fed to the inlet distributor which is mounted downstream of the lower electrode.1 Both the distributor and the lower electrode contain 126 holes 1mm in diameter each. The bottom of the distributor acts as an exit grid for the plasma-precursor mixture. The power injected in the discharge can be set between 60W and 100 W. All the experiments have been carried out a) at a fixed power of 80W.

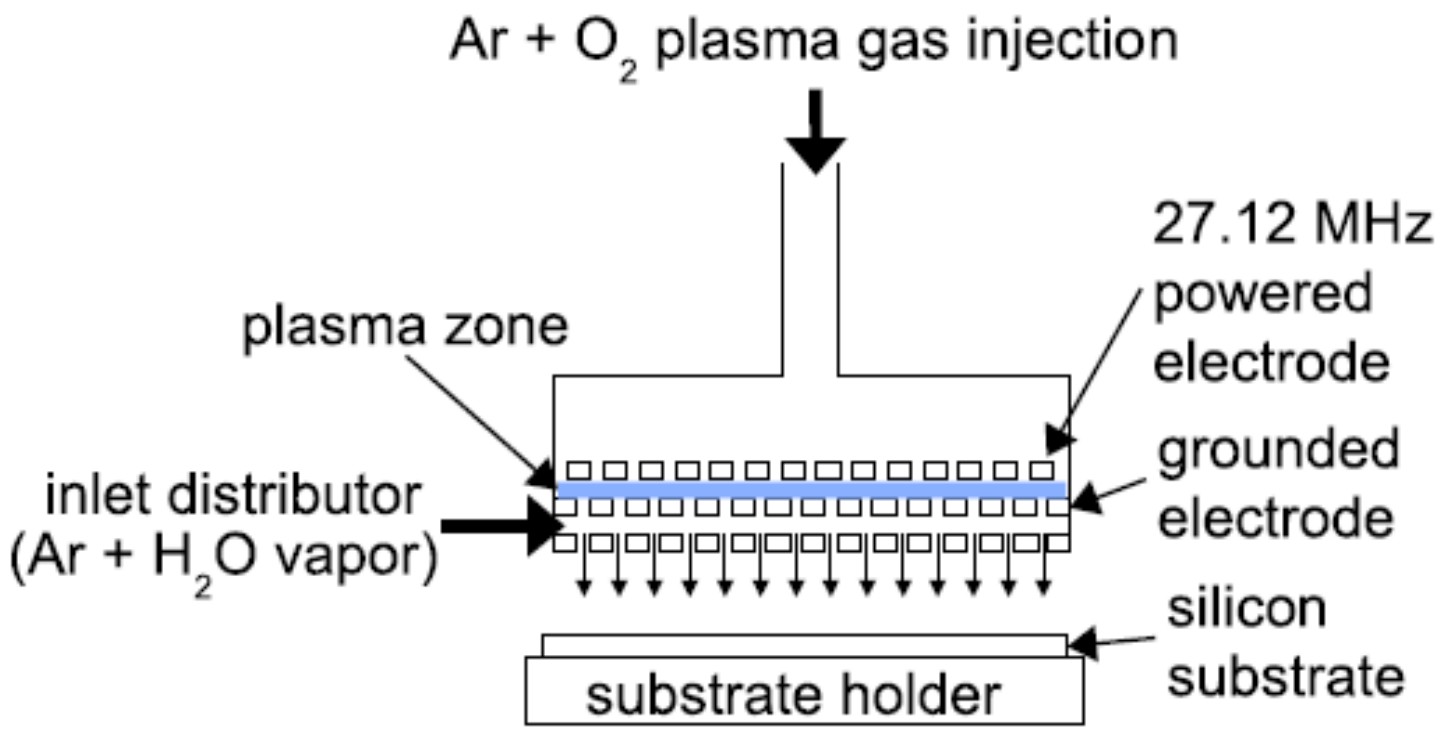

***FIG. 1. Schematic of the plasma torch and substrate holder.***

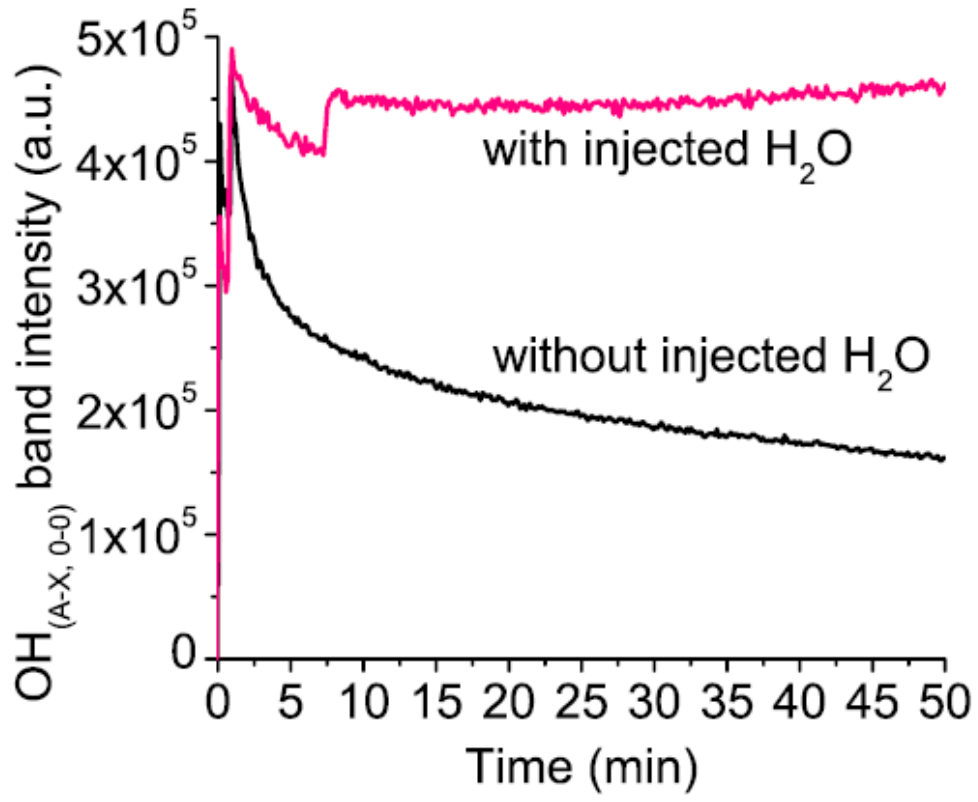

***FIG. 2. OH ($A^2\Sigma^+$; v'=0 → $X^2\Pi$; v"=0) band integrated intensity versus plasma on time, with and without injected water vapor (Ar 30 L/min, $O_2$ 0 mL/min, 80 W, gap= 5 mm).***

A silicon substrate, cut to a rectangle of dimensions 70mm*80 mm, is placed downstream of the plasma torch, at a distance varying from 2mm to 13 mm. The torch-to-substrate separation distance will also be denoted as "gap" in some figures, the two terms being used interchangeably. The surface of the silicon substrate was previously etched in a $SF_6/O_2$ plasma and appears black due to the presence of columnar microstructures;10 this treatment was done to decrease the reflectivity of the sample and limit the detection of stray light from the afterglow.

OH radicals are easily detected in our conditions by optical emission spectroscopy (OES) through the $A^2\Sigma^+$,v'=0→$X^2\Pi$,v"=0 emission band around 309 nm. Ar and $O_2$ gas purities are 99.999%, and several tests lead us to conclude that the OH production mostly comes from water that adsorbs on the inner walls of the plasma torch when the plasma is turned off. The evolution of the emission intensity from excited OH ($A^2\Sigma^+$), as shown in Figure 2, suggests that the OH concentration decays as the plasma on time increases. Therefore, for all the experiments presented in this study, water vapor is injected via the inlet distributor to allow for a better stabilization of the OH concentration. The water vapor mass flow rate carried by Ar in the bubbling system is approximately 9mg/min, determined from mass measurements of the bubbler. For a total Ar flow rate of 30 L/min, this represents a concentration of water molecules of 370 molar ppm. When no oxygen is injected in the plasma torch, one $H_2O$ molecule can give at most one OH molecule through dissociation, so the injected water vapor can supply a maximum OH radical density of 5-9.10$^{15}$ cm$^{-3}$ for gas temperatures in the range 300–460 K.

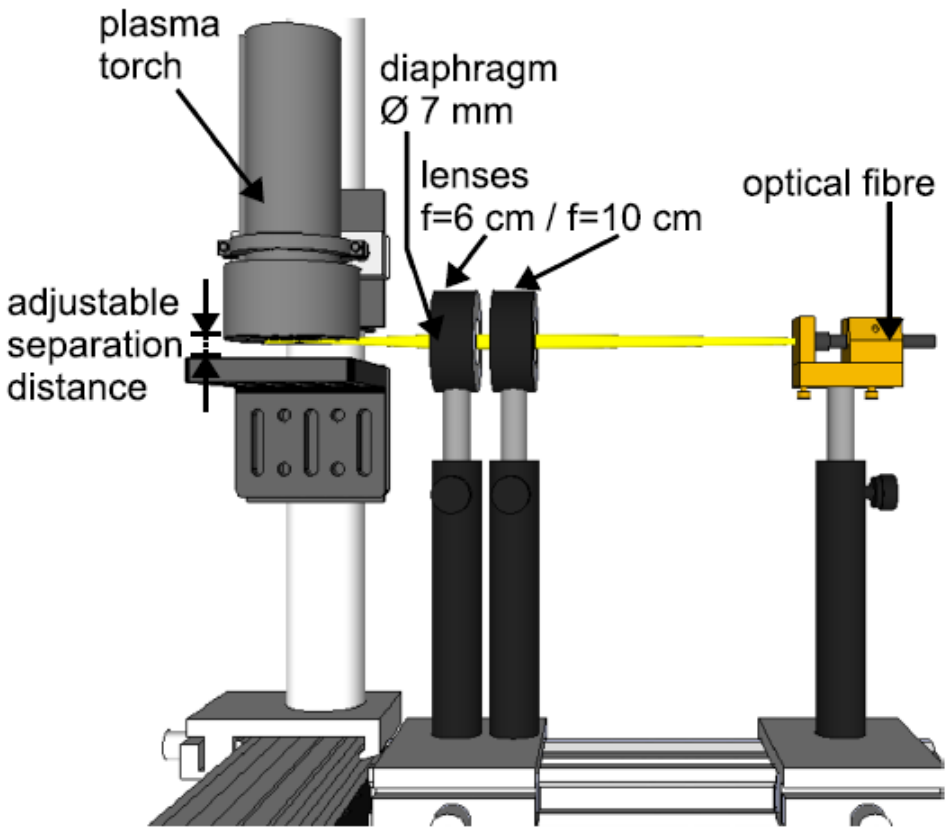

***FIG. 3. Experimental set-up for optical emission spectroscopy in the plasma afterglow.***

## 2.2. Optical emission spectroscopy

OES has been performed with a Spectra Pro 2500i Acton Research spectrometer equipped with a CCD camera (Princeton digital camera with 400*1340 pixels). As depicted in Figure 3, light emitted by the plasma afterglow is focused by a set of two converging lenses into an optical fiber and sent to the entrance slit of the spectrometer. Optical emission spectra in the UV range are acquired with a 2400 grooves/mm grating blazed at 240 nm (linear dispersion of 0.745 nm/mm at a wavelength of 254 nm). The width of the entrance slit being set to 10 lm, the FWHM of the $Hg_{254nm}$ from a mercury lamp is 3.6*10$^{-2}$ nm. A 1800 grooves/mm grating blazed at 500 nm (linear dispersion of 0.846 nm/mm at a wavelength of 579 nm) is used for optical emission spectra in the visible–near infrared (VNIR) range. With the same setting of the entrance slit, the FWHM of the $Hg_{579nm}$ line is 4.2*10$^{-2}$ nm. For measurements in the VNIR range, a yellow glass high-pass filter is inserted just before the optical fiber to eliminate the second order of OH and $N_2$ band emission.

Absolute densities of Ar excited atomic states have been determined experimentally through absolute line intensity measurements. Given the size of the plasma imaged, we assumed the plasma to be optically thin, i.e., possible absorption by the plasma is neglected. The captured flux $\Phi$ (W) integrated on a transition u→l is given by

$$\phi = \frac{\Omega}{4\pi} V T N_u A_{ul} h\nu_{ul},$$

where $\Omega$/=4$\pi$ is the fraction of solid angle detected, V the emitting plasma volume (cm$^3$), T the transmission of the optical system, Nu the density of atoms in the upper state (cm$^{-3}$), $A_{ul}$ the Einstein coefficient for spontaneous emission (s$^{-1}$), and $h\nu_{ul}$ the energy of the transition (J).

The emitting plasma volume and solid angle of detection have been determined by injecting white light backwards from the end of the optical fiber and measuring the dimensions of the cone imaged at the position of the plasma afterglow. A diaphragm, 7mm in diameter, is located just before the lens of collection, to increase the depth of field and reduce the solid angle of detection down to 0.01 sr. The focal point, i.e., the image of the optical fiber entrance, is 1mm in diameter. The emitting plasma volume is then approximated to be a cylinder of diameter 1mm and length 16 mm, corresponding to sixteen holes of the lower electrode in the line of sight. All the OES measurements have been carried out with the line of sight centered relative to the circular plasma torch and at an axial position centered at 0.5mm (the origin is the exit grid of the plasma torch). Screening by the plasma torch and/or the substrate holder is accounted for by calculating the ratio of the shadowed cross section to the full cross section of the cone at a known distance from the lens and dividing the solid angle value by this ratio. In most situations, the solid angle is reduced to about 70%. The transmittance of the lenses is assumed to be 96% per dioptre, hence 84.9% for the two lenses. The transmittance of the high-pass filter has been determined experimentally. The signal intensity measured by the CCD camera has been calibrated to an equivalent flux by using an irradiance calibrated tungsten-halogen lamp positioned at a distance of 50 cm from the optical fiber (without any optics). The flux emitted by the lamp is calculated by multiplying the nominal irradiance to the surface area of the optical fiber (1.22 $mm^2$).

## 2.3. Ozone concentration measurements by absorption of mercury emission line

An estimation of the ozone concentration in the plasma afterglow has been obtained by measuring the absorption of the $Hg_{253,65nm}$ emission line from a mercury lamp. The mercury lamp and optical fiber are positioned at opposite sides of the plasma torch. Pinholes of diameter 1mm are inserted right after the lamp and before the optical fiber to limit the detection of stray light. The absorption path length l is about 10 cm. Assuming a homogeneous concentration, the ozone concentration $n_{O3}$ is determined from the measurement of $I_0$, the intensity of mercury line with the plasma off, and I, when the plasma is turned on, according to the Beer-Lambert law,

$$n_{O_3} = \frac{1}{\sigma l} \ln\left(\frac{I_0}{I}\right)$$

where $\sigma$ is the absorption cross section for ozone, which equals $1.15*10^{-17}$ $cm^{-2}$ at 254 nm.11 The absorption cross sections for $O_2$, $N_2$, and $H_2O$ at this wavelength are $<5*10^{-24}$ $cm^2$, 12,13 $<5*10^{-24}$ $cm^2$,12 and $<9*10^{-22}$ $cm^2$, 14 respectively. The $Hg_{253,65nm}$ emission line is not resonant with OH absorption lines with rotational levels lower than N"=15. 15 The influence of absorption by molecular species other than $O_3$ is therefore neglected.

In our experimental conditions, other diagnostics employed suggest that the concentration of ozone is not constant along the absorption path length, so the values obtained should be considered as indicative only. To account for slow temporal variations in intensity of the mercury lamp, four series of measurements of I then $I_0$ have been taken by switching alternatively on and off the plasma. As a result of the uncertainty on $I=I_0$ and the short absorption path length, the lowest detectable number density of $O_3$ molecules is $10^{14}$ $cm^{-3}$.

## 2.4. Mass spectrometry

A quadrupolar mass spectrometer (Balzers Omnistar 300) adapted to atmospheric pressure measurements has been employed to get information on relative concentrations of major stable species in the plasma afterglow. The ionization chamber of the mass spectrometer is evacuated to a pressure of $1\text{-}4*10^{-7}$ mbar by a turbomolecular pump assisted by a primary pump. A capillary tube, 100 lm in internal diameter and about 1m long, is used to probe neutral species, the tip of which is positioned horizontally in between the plasma torch and the substrate. To obtain maps of concentrations, the plasma torch system is moved relative to the capillary both horizontally and vertically.

For our experiments, the ionization energy is set to 70 eV. The dynamic range of measurements is $10^4$, hence only species with concentrations higher than 100 ppm can be detected. Ar, $N_2$, $O_2$, $CO_2$, and $H_2O$ species have been considered for the analysis.
The relative concentration xi of each stable species i in the plasma afterglow is calculated using the relationship,

$$x_i = \frac{(I_{mp_i} - BG_{mp_i})f_i}{\sum_i (I_{mp_i} - BG_{mp_i})f_i}$$

where $I_{mpi}$ is the intensity of the main ion peak for species I and $BG_{mpi}$ its background intensity. Whatever the plasma conditions and the plasma torch-to-substrate separation distance, a residual background containing $N_2$, $O_2$, $CO_2$, $H_2O$, F, and Cl is always detected. By inserting the capillary inside a tube of flowing argon (purity 99.999%), this background signal can be determined and subsequently subtracted.
$f_i$ is a correction factor that allows measurements of the main ion peaks to account for isotopes and fragmentation into ions of lower mass. It is determined experimentally for each species i from the mass spectrum of air and is equal to

$$f_i = 1 + \frac{\sum_k I_{sp_{ik}}}{I_{mp_i}}$$

with $I_{spik}$ the intensity of a secondary ion peak for species i. Table I lists the m/z values of the main and secondary ion peaks for each species considered in the present analysis, along with the experimentally calculated values of the factors $f_i$.

| Species | Main peak *m*/*z* | Secondary peaks *m*/*z* | $f_i$ |
|---|---|---|---|
| $H_2O$ | 18 | 17 | 1.25 |
| $N_2$ | 28 | 29, 30, 14 | 1.08 |
| $O_2$ | 32 | 34, 16 | 1.11 |
| Ar | 40 | 36, 20 | 1.23 |
| $CO_2$ | 44 | 12 | 1.12 |

***TABLE I. Factors $f_i$ used for the determination of Ar, N2, $O_2$, $CO_2$, and $H_2O$ relative concentrations from mass spectrometry measurements.***

## 2.5. Laser induced fluorescence on OH radicals

In the scope of this study, LIF experiments on OH radicals have been carried out to gain insight into their collisional environment. Figure 4 shows the schematic of the experimental set-up for the LIF diagnostic of the plasma afterglow. The OH(A–X, 0–0) band excitation around 309 nm has been obtained with a Sirah Cobra-Stretch dye laser pumped by a pulsed Nd:YAG laser. The frequency doubled emission at 532 nm of the Nd:YAG laser pumps a DCM dye with methanol mixture to provide emission at around 618 nm. A second harmonic generator (denoted SHG in Figure 4) is then used to obtain the wavelength of 309 nm. The laser pulse duration is about 8 ns, the repetition rate 10 Hz, and the linewidth about $8*10^{-4}$ nm (constructor data).

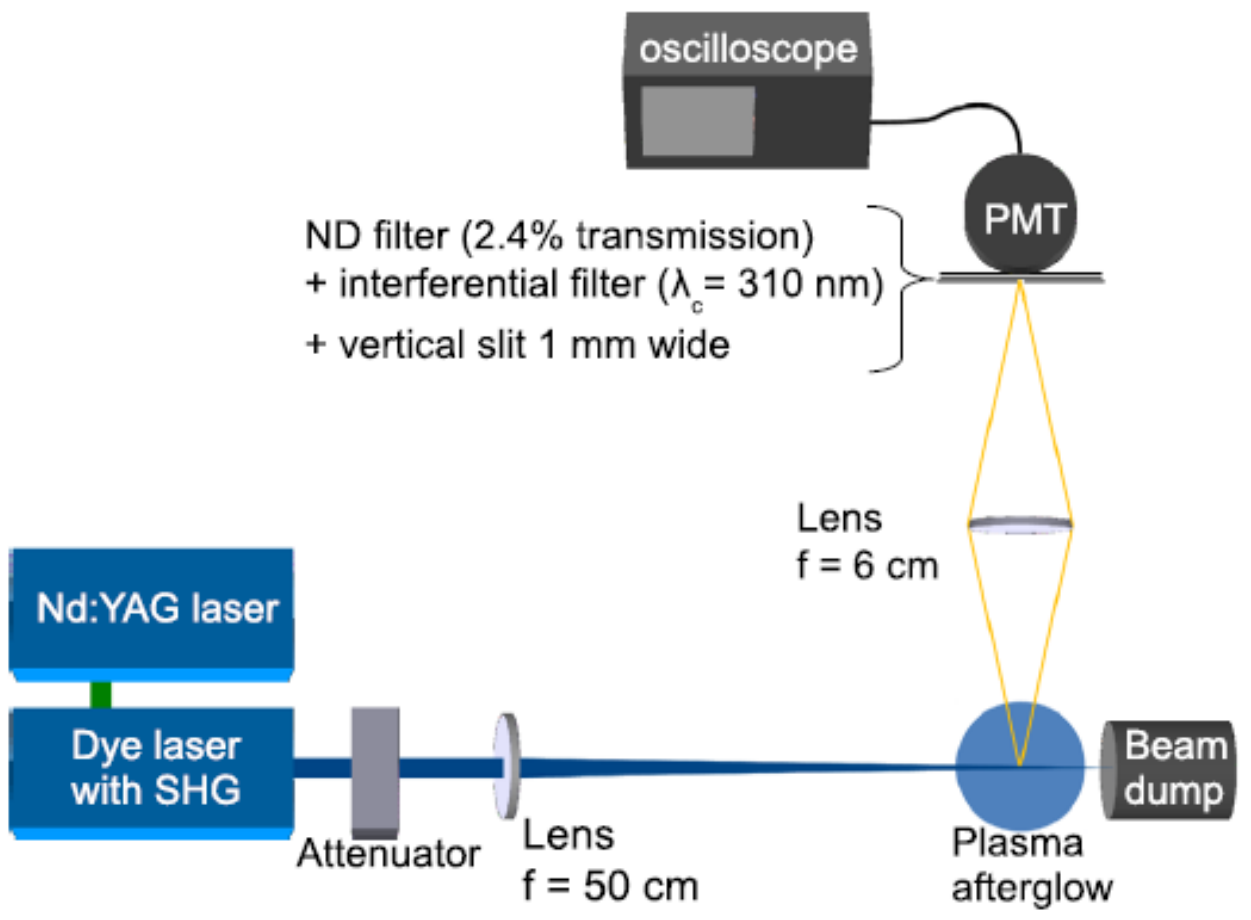

***FIG. 4. Schematic of the experimental set-up for LIF measurements on OH radicals in the plasma afterglow.***

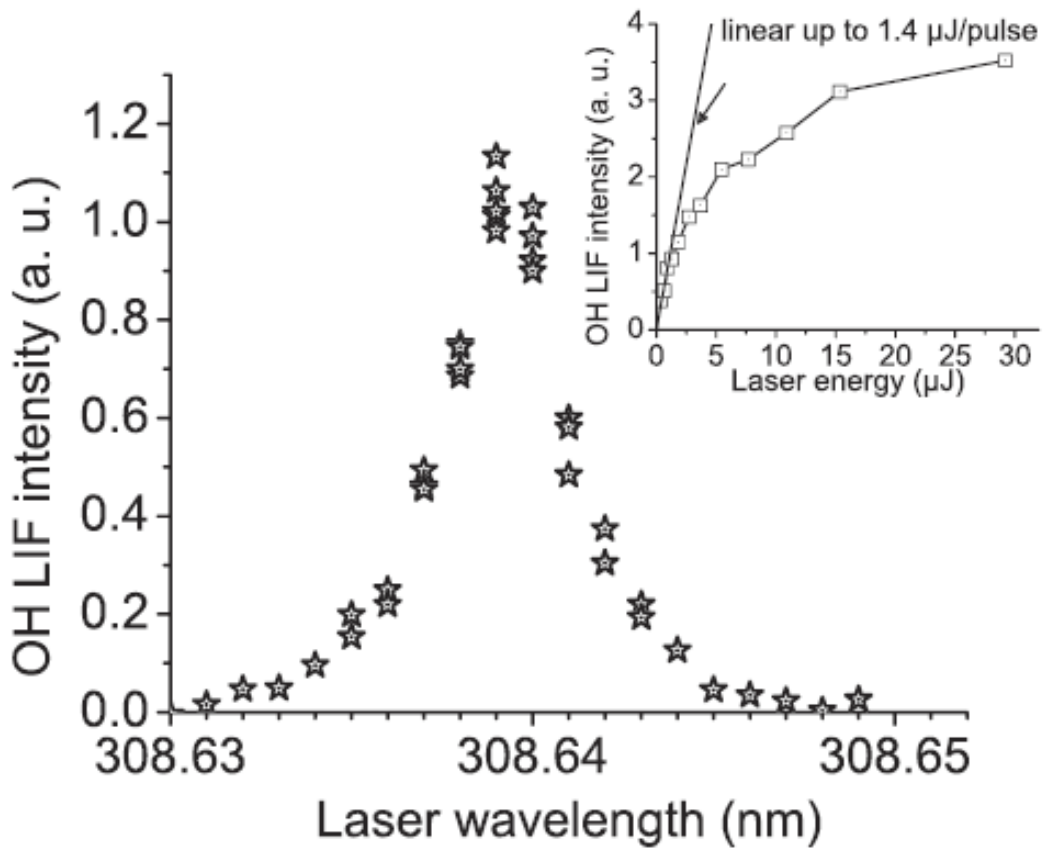

***FIG. 5. OH LIF excitation scan on P1(2) transition (laser energy 1.3 µJ/pulse); inset: LIF signal at the center of the P1(2) transition versus laser energy to determine the range of linearity (Ar 30 L/min, $O_2$ 0 mL/min, 80 W, gap 5 mm).***

The laser beam energy can be varied from 0.3 to 2 mJ/pulse. It is attenuated by a series of beamsplitters and by one or several neutral density (ND) filters before the beam is focused into the plasma afterglow by a fused silica spherical lens of focal length 50 cm. A horizontal slit of width 2mm is inserted in front of the lens to cut 2/3 of the beam height; the laser spot at the entrance of the plasma afterglow region is 0.8mm high and 0.5mm wide.

The fluorescence emitted by the OH radicals is collected by a fused silica spherical lens of focal length 6 cm and sent to a photomultiplier tube (PMT Hamamatsu R928) after passing trough an assembly comprised of a vertical slit of width 1mm, a narrowband interferential filter (center at 310 nm, FWHM 10nm) and a ND filter to prevent the PMT from saturating. The lens and vertical slit are positioned such that the magnification is equal to 1. Therefore, the plasma volume sampled by LIF is approximately 1mm*0.5mm*0.8mm. The signal detected by the PMT is recorded on an oscilloscope (Tektronics DPO3032 with a bandwidth of 300MHz and a sample rate of 2.5 Gs/s) and averaged over 128 laser pulses to reduce laser energy fluctuations.

The laser pumping is set on the P1(2) transition at 308.639 nm which excites the F1(1) level of the upper OH($A^2\Sigma^+$; v'=0) vibrational state from the f1(2) level of the ground OH($X^2P$; v"=0) vibrational state. Other levels in the v'=0 state may be populated by collision induced rotational energy transfer (RET). Broadband fluorescence on the (0–0) band is then detected. The choice of the pumping wavelength was motivated by the low lying energy of the pumped level; hence, it is one of the highest populated levels in the $X^2\Pi$; v"=0 state. Furthermore, the transition is isolated and the wavelength is in the acceptable range of the laser dye. Figure 5 shows the excitation scan on the P1(2) transition for a laser energy of 1.3 µJ/pulse, and the time-integrated LIF signal intensity at the center of the P1(2) transition versus the laser pulse energy. A linear relationship is verified for laser energy up to 1.4 µJ/pulse, therefore, all the experiments have been carried out at an energy around 1.3 µJ/pulse.

The v"=0→v'=0 excitation scheme has been widely used for OH detection in flames16–18 and in the troposphere.19 Since the method employs the same bands for excitation and detection, it can induce experimental uncertainties due to Rayleigh scattering of the laser beam and possible reabsorption of fluorescence. Nonetheless, the excited vibrational state can only decay through radiative de-excitation or collision induced electronic quenching, which makes it easier than schemes with excitation of v′=1 (Refs. 20–22) or v′=3 (Refs. 23 and 24) (where the excited vibrational state can also decay through collision to lower lying vibrational states in $A^2\Sigma^+$ before fluorescence occurs) for measuring the quenching rate and estimating the concentration of collision partners.

# 3. Results

## 3.1. Plasma emission

The absolute densities of Ar excited states have been determined close to the exit grid of the plasma torch (axial position 0.5 mm, with a spatial resolution of 1 mm) through OES measurements, for a varying torch-to-substrate separation distance. Most of the detected emission is in the 690-1050 nm range, corresponding to transitions that originate from Ar 4p states (12.5–13.5 eV). Minor emission lines are detected from higher excited states, namely, the 4d (14.7 eV) and the 5p states (14.5 eV), and no ionic lines are visible.

An Ar excitation temperature, $T_{13}$, is then derived from the calculation of the Ar ground state density (level 1) and from the measured Ar 4p states density (denoted as level 3, the level 2 corresponding to the Ar 4s atomic states, which are either metastable states or resonant states emitting in the vacuum ultraviolet region). Because of the energy difference between the ground state and the 4 p states, this method is more accurate than comparing only excited state densities to obtain an excitation temperature in non-equilibrium plasmas.25 A collisional radiative model is then needed to convert the estimated temperature $T_{13}$ to an electron temperature, which is beyond the scope of this study. Variations in $T_{13}$ only are presented and give a measure of the evolution of the whole Ar 4 p population. The Ar ground state density n1 is obtained from the ideal gas law, $P=n_1kT_{gas}$, where P is 1 bar and $T_{gas}$ is estimated by the OH($A^2\Sigma^+$; v′=0) rotational temperature. This equation is used assuming that the population of the Ar ground state is much greater than that of any other species in the plasma afterglow. To obtain the OH(A) rotational temperature, the OH (A–X, 0–0) band emission is fitted to simulated spectra using the software LIFBASE.15 For plasmas ignited in Ar with added water vapor, Trot is always found in the 400–450K range.

The use of excited OH(A) rotational temperature to estimate the gas temperature in plasmas has been discussed in various publications.26–29 In order to assume that both temperatures are equal, the OH(A) rotational population must follow a Boltzmann distribution at the translational temperature of neutral species, i.e., a thermalized distribution. This is only possible if the OH($A^2\Sigma^+$; v′=0) state is directly created with a thermalized rotational population distribution or if the state can relax to a thermalized one by RET before deexcitation by emission or electronic quenching occurs. OH(A) radicals can be produced efficiently by collisions of $H_2O$ molecules with metastable Ar atoms.30,31 A three-step dissociative excitation mechanism is proposed in Ref. 31, where OH(X), and OH(A) radicals at high rotational levels (with a maximum at N′=10–11) are the main products. Dissociative excitation of $H_2O$ molecules by electron impact is another channel to produce OH(A) radicals, this requires electrons of energies greater than 9 eV.32 As it will be detailed in Sec. IIID, RET through collisions with argon atoms is very efficient, and for most cases the effective lifetime of OH($A^2\Sigma^+$; v′=0) state is greater than the RET time constant by at least an order of magnitude, so we will consider that the state is in thermal equilibrium. A further check on the rotationally resolved OH(A–X, 0–0) emission spectrum indeed reveals a slight overpopulation of rotational levels N′=10–18. However, their overall population does not exceed 3%, and the v2 error on the fit is not significantly improved by changing from a pure

Boltzmann distribution to a modified distribution. Therefore, the OH(A) rotational temperature should provide a good estimation of the gas temperature, $T_{gas}$, close to the exit grid of the plasma torch.

The inset in Figure 6 illustrates the determination of $T_{13}$ from a semi-logarithmic plot of N/g, the argon electronic state populations divided by the respective level degeneracies, versus the level energies E. The slope of the linear regression between the Ar 4s state and the 4p states is $-1/kT_{13}$ (k is the Boltzmann constant). The uncertainty on $T_{13}$, reported as error bars on the values of Figure 6, is calculated from the standard deviation of the slope of the linear regression. It should be noted that the 4p state densities measured are averages over the line of sight.

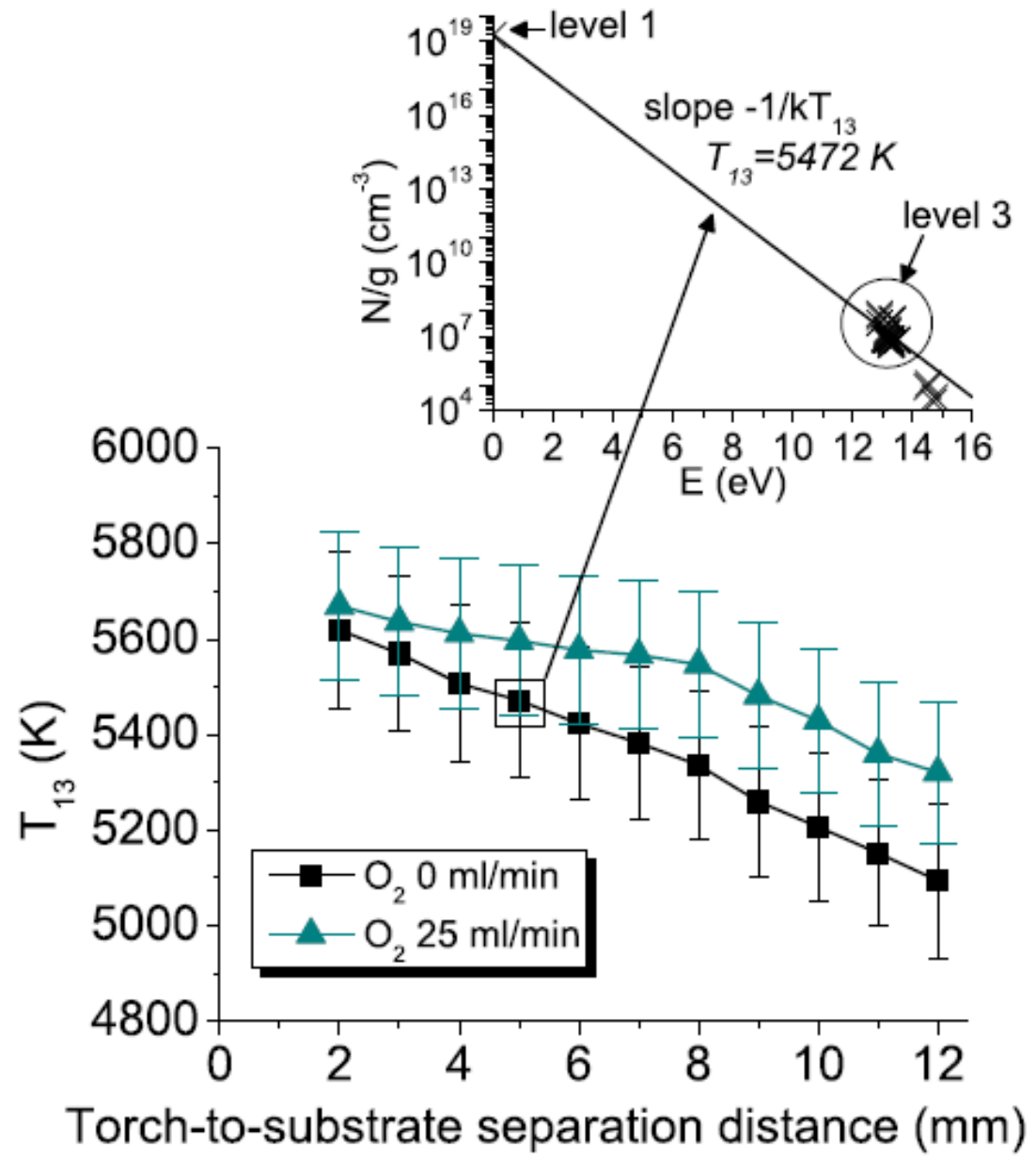

***FIG. 6. Ar excitation temperature $T_{13}$ versus plasma torch-to-substrate separation distance (Ar 30 L/min, 80 W); inset: graphic determination of $T_{13}$.***

Calculated excitation temperatures $T_{13}$ are in the range 5000–5800 K, the values being always greater when $O_2$ is added to Ar. Most interestingly, a decrease in $T_{13}$ is observed as the plasma torch-to-substrate distance increases, with and without $O_2$ in the gas mixture. Figure 7(a) indicates that a significant emission from the second positive system of $N_2$ is measured for plasma torch-to-substrate separation distances greater than 6mm. The emissivity of the $N_2$ $C^3\Pi^u \rightarrow B^3\Pi^g$ (0–2) transition around 380 nm is seen to increase with the plasma torch-to-substrate separation distance up to a distance of 9 mm, from which a decay is observed. The more intense $N_2$ (C–B, 0–0) transition around 337 nm is also detected (cf. Figure 7(b)), but the emission band overlaps another molecular band which is attributed to the NH($A^3\Pi \rightarrow X^3\Sigma^-$) system.33,34 Emission from NH(A) species is detected for shorter torch-to-substrate separation distances than for $N_2$(C) species, which suggests that if $N_2$ molecules diffuse into the afterglow, they are completely dissociated, or that $N_2$ impurities are flowing with the injected Ar with water vapor mixture. Since the NH(A–X) emission intensity is much lower than that of $N_2$(C–B) for a gap of 9mm and since no emission from $N_2$(C) species is detected for small gaps, we can hypothesize that the air diffusion is weak in the range 2–5 mm. Evidence of this hypothesis will be given in Secs. III C and IIID.

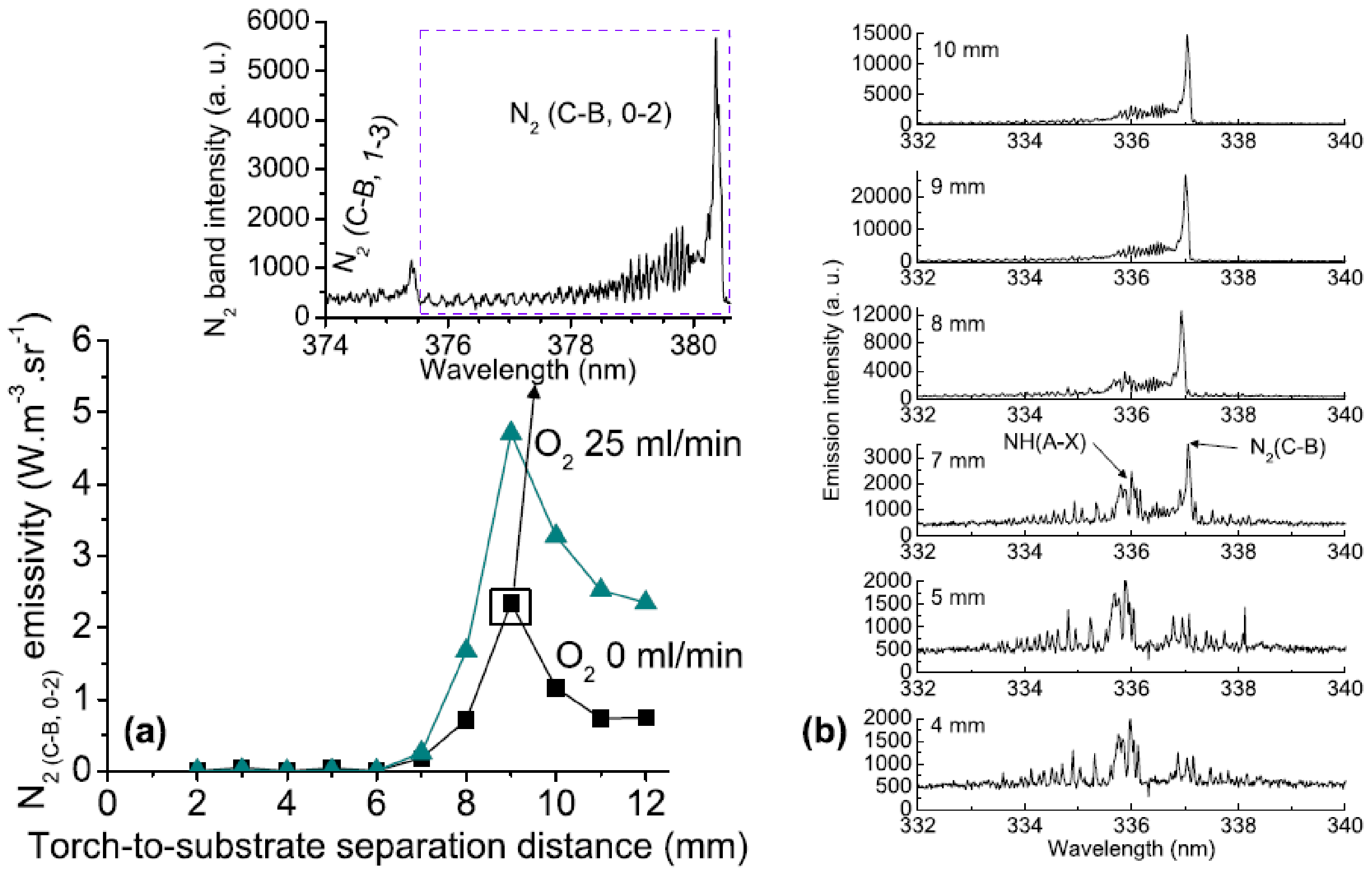

*FIG. 7. (a) $N_2$ $C^3\Pi_u \rightarrow B^3\Pi_g$ (0–2) band integrated emission coefficient versus plasma torch-to-substrate separation distance (Ar 30 L/min, 80 W)—inset: $N_2$ (C–B) band emission spectrum showing the (1–3) and (0–2) transitions (Ar 30 L/min, $O_2$ 0 mL/min, 80 W, gap 9 mm); (b) emission spectra in the 332–340nm wavelength range for various plasma torch-to-substrate separation distances (Ar 30 L/min, $O_2$ 0 mL/min, 80 W).*

## 3.2. Ozone concentration

Measured, line averaged, ozone concentrations close to the exit grid of the plasma torch and to the substrate are shown in Figure 8(a) as a function of the plasma torch-to-substrate separation distance. The $O_2$ flow rate injected is 25mL/min. The lowest detectable concentration is $1.10^{14}$ $cm^{-3}$, therefore, the ozone concentration is significantly higher than this limit only for plasma torch-to-substrate separation distances greater than 8mm. It then increases as the substrate is moved further away from the plasma torch.

The evolution of ozone concentration versus the oxygen flow rate is displayed in Figure 8(b) for a gap of 10 mm; measurements were carried out at axial positions 2mm and 9 mm. The ozone concentration is seen to increase with the increase in oxygen flow rate. Graphs (a) and (b) of Figure 8 indicate that it is both necessary to inject some oxygen in the plasma torch and to move the substrate at separation distances greater than 8mm to produce ozone at concentrations higher than the detectable limit. It is also noticeable that the concentrations of ozone measured close to the substrate are almost equal to those measured close to the exit grid of the plasma torch.

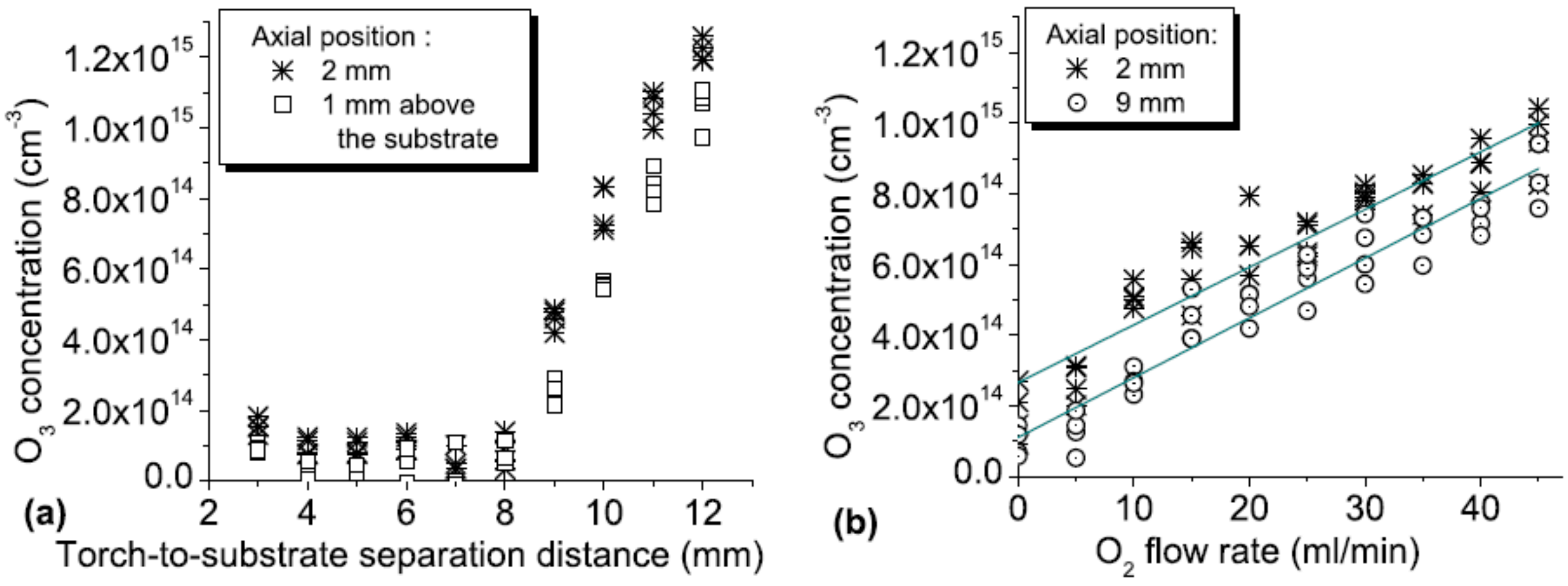

*FIG. 8. Average ozone concentration versus (a) the plasma torch-to-substrate separation distance (Ar 30 L/min, $O_2$ 25 mL/min 80 W) (b) the $O_2$ flow rate (Ar 30 L/min, 80 W, plasma torch-to-substrate separation distance 10 mm).*

## 3.3. Ar, $N_2$, and $O_2$ relative concentrations by mass spectrometry

Figure 9(a) presents Ar, $O_2$, and $N_2$ relative concentrations obtained from mass spectrometry measurements ($H_2O$ and $CO_2$ molecules were included in the total composition) versus plasma torch-to-substrate separation distance at a radial position of 0mm and an axial distance of 2mm. Positioning of the metallic capillary closer to the exit grid of the plasma torch causes plasma instabilities with arcing effects. At this location, $N_2$ and $O_2$ molecules are only detected for gaps greater than 7 mm, and their concentration dramatically increases as the substrate is moved further away from the torch.

In Figure 9(b), the evolution of $N_2$ relative concentration is represented for several radial positions of measurement. This graph highlights the presence of radial gradients in $N_2$ concentration for gaps larger than 6 mm. Such an effect is further illustrated in Figure 10 where the $N_2$ relative concentration is plotted against the axial and radial position, for gaps of 4mm, 5mm, 7 mm, and 10 mm. For a gap of 4mm, the N2 concentration is not significant, and for a gap of 5mm, $N_2$ (and $O_2$) molecules start to be detected close to the substrate at a radial position of 12.5 mm, their concentration reaching 2%. Then for gaps of 7mm and 10 mm, radial and axial gradients in $N_2$ (and $O_2$) are measured. For a gap of 10 mm, the $N_2$ concentration reaches 72% at a radial position of 12.5mm and an axial position of 2 mm.

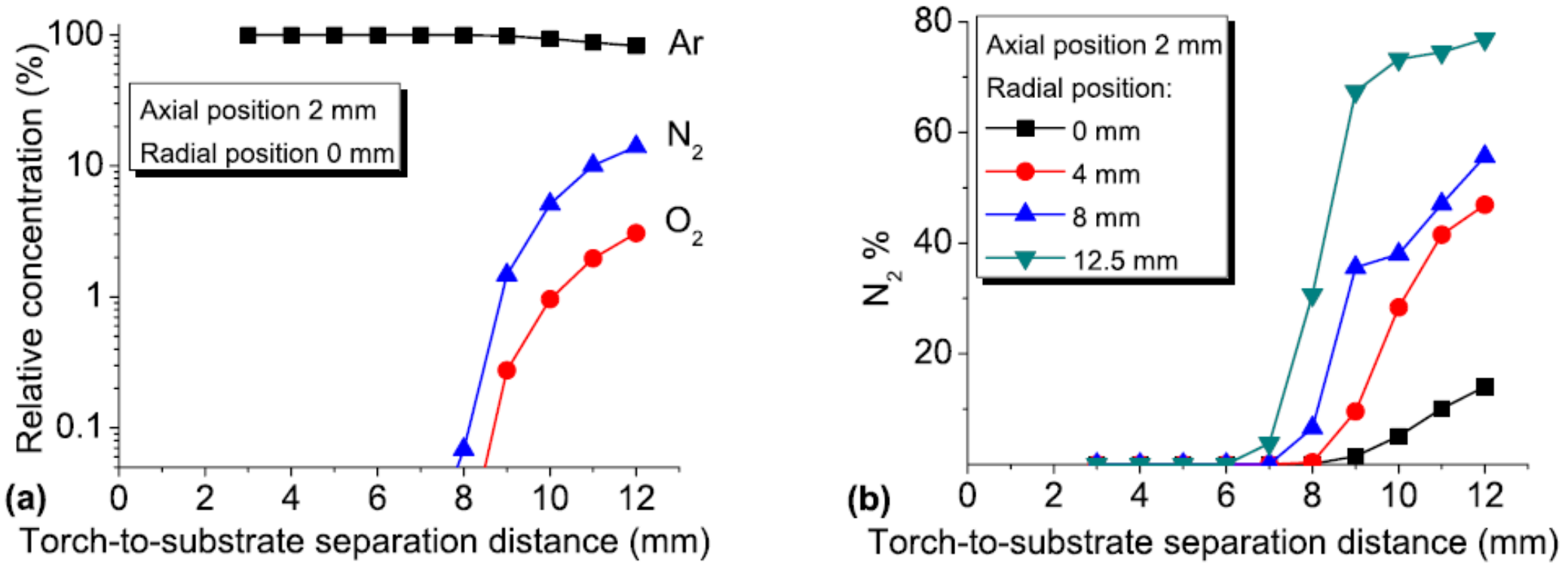

*FIG. 9. Relative concentrations determined from mass spectrometry measurements at the axial position of 2mm (Ar 30 L/min, $O_2$ 0 mL/min, 80 W): (a) Ar, $N_2$, and $O_2$ relative concentrations versus plasma torch-to-substrate separation distance at a radial position of 0 mm; (b) $N_2$ relative concentration versus plasma torch-to-substrate separation distance at various radial positions.*

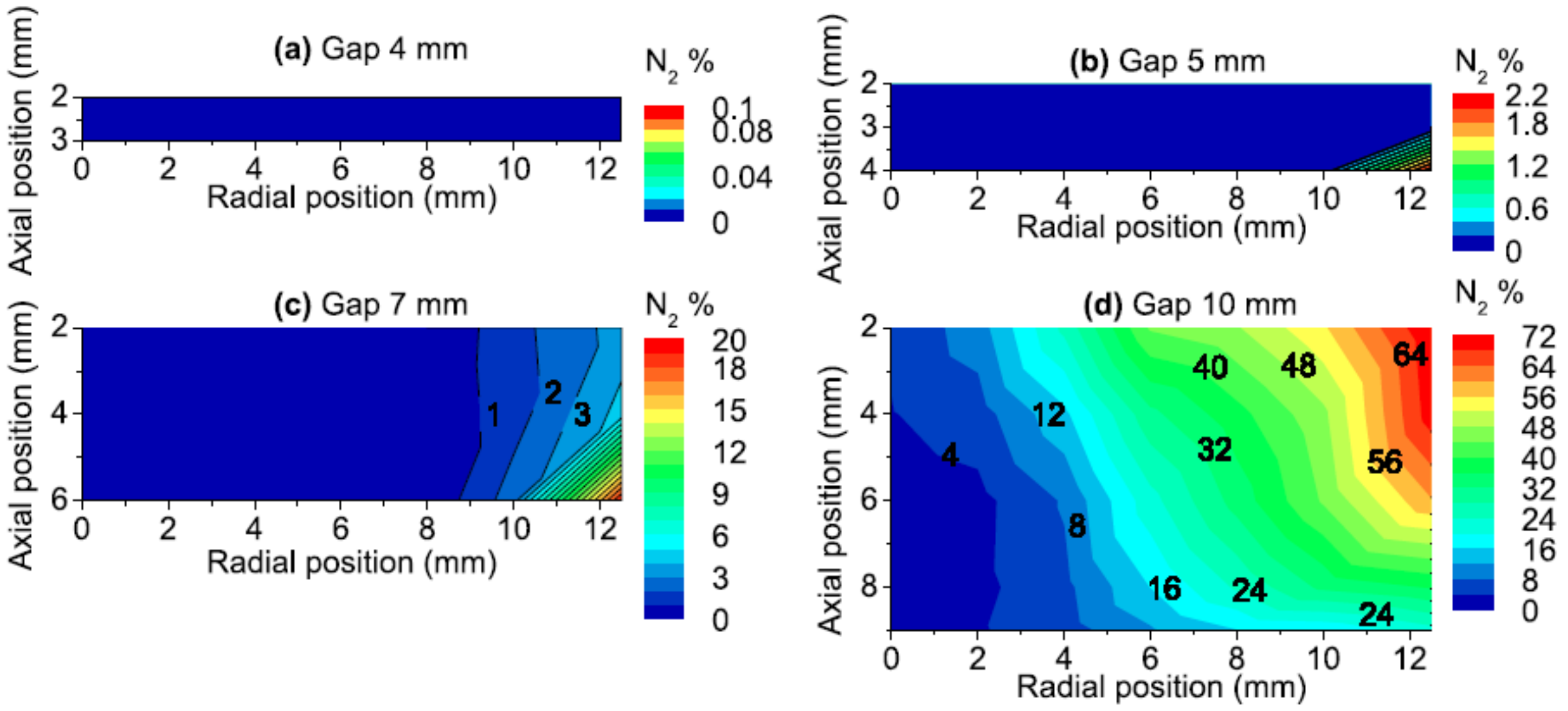

***FIG. 10. $N_2$ relative concentration determined from mass spectrometry measurements versus axial and radial position in the plasma afterglow (Ar 30 L/min, $O_2$ 0 mL/min, 80 W) for a plasma torch-to-substrate separation distance of (a) 4 mm, (b) 5 mm, (c) 7 mm, and (d) 10 mm.***

## 3.4. Laser induced fluorescence on OH radicals

The air content under the plasma torch has been estimated by analyzing the decay rate of OH radicals excited by laser pumping to the F1(1) level of the $A^2\Sigma^+$; v'=0 vibrational state. The temporal evolution of the broadband measured LIF signal $I_{LIF}$ is fitted to a single exponential decay of the form,

$$I_{LIF}(t) = a \cdot e^{-t/\tau_{eff}}$$

If the upper vibrational state achieves thermal equilibrium through RET before de-excitation, the effective lifetime $\tau_{eff}$ of the LIF signal corresponds to

$$\tau_{eff}^{-1} = A + Q = A + \sum k_x n_x$$

with A the Einstein coefficient for spontaneous emission, Q the total quenching rate, $k_x$ the rate constant for collisional quenching of OH by collider x and $n_x$ the concentration of collider x. A, Q, and $k_x$ are averaged over the whole v'=0 vibrational state.
The time constant for rotational relaxation is calculated to be in the range 0.1–0.4 ns according to Refs. 35 and 36. Wysong et al. estimated that laser pumping of OH radicals, initially in the $X^2\Pi$ v"=0 state at room temperature, to the N'=2 and N'=5 levels of the $A^2\Sigma^+$ v'=0 state, resulted in a thermalized rotational population distribution in 5 Torr of argon after 50 ns and 300 ns, respectively.37 Extrapolating to atmospheric pressure, thermalization out of the level N'=1 should then be complete in a maximum time of 2 ns. The effective lifetime values are determined by fitting the temporal LIF signal to an exponential decay from 10 ns onwards after the end of the laser pulse.

From the knowledge of the rate constants $k_x$ and the Einstein coefficient A, it is possible to obtain information on the concentrations of collision partners. A is taken as $1.45*10^6$ $s^{-1}$ from Ref. 15. Table II shows the rate constants for collisional quenching of OH($A^2\Sigma^+$; v'=0) by several species at a temperature of 400K (cf. Sec. IIIA).

| Species | Rate constant $k_x$ $cm^3\,s^{-1}$ | Reference |
|---|---|---|
| $N_2$ | $1.70 \times 10^{-11}$ | 18 |
| $O_2$ | $1.28 \times 10^{-10}$ | 18 |
| $CO_2$ | $3.10 \times 10^{-10}$ | 18 |
| $H_2O$ | $5.82 \times 10^{-10}$ | 18 |
| Ar | $8.8 \times 10^{-14}$ | 30 |
| Air | $4.66 \times 10^{-11}$ | Calculated |

***TABLE II. Rate constants for collision induced electronic quenching of OH ($A^2\Sigma^+$; v'=0) by several species at 400 K.***

The gas composition is determined from a mixture of species coming from the plasma torch and air species. Ar and $H_2O$ are taken into account for the plasma gas mixture. The air composition is chosen as $N_2$ 77.18%, $O_2$ 20.71%, $H_2O$ 1.15%, Ar 0.92%, $CO_2$ 380 ppm (air composition at 1013 hPa, relative humidity of 50%).

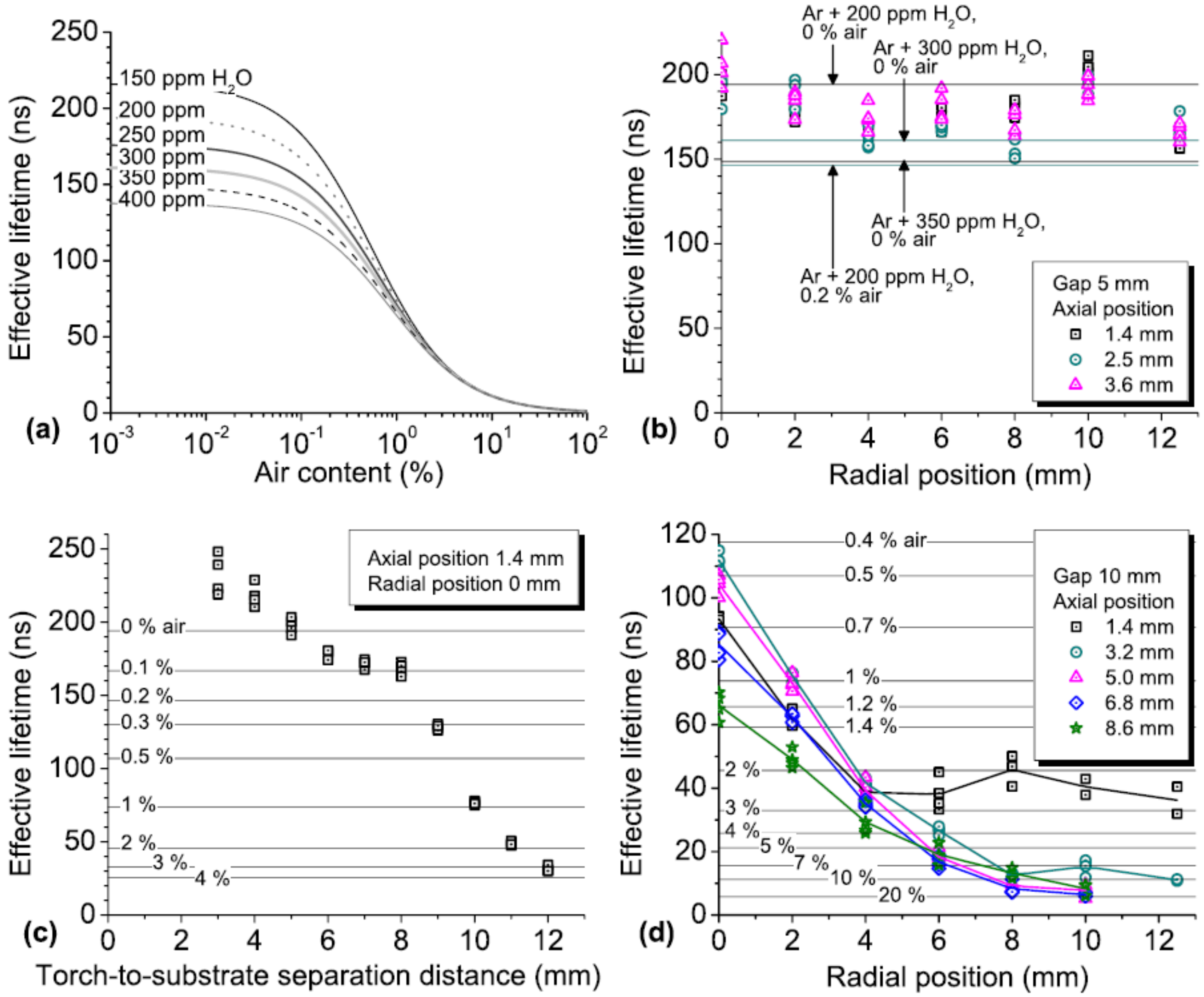

***FIG. 11. (a) Simulated effective lifetime of OH LIF signal versus air content for various H2O contents in the plasma gas mixture (Ar 30 L/min, $O_2$ 0 mL/min, $H_2O$ 150–400 ppm, air with 50% relative humidity); experimental effective lifetime of OH LIF signal (Ar 30 L/min, $O_2$ 0 mL/min, 80 W) (b) versus radial position for various axial positions at a fixed gap of 5 mm, (c) versus plasma torch-to-substrate separation distance (axial position 1.4 mm, radial position 0 mm), (d) versus radial position for various axial positions at a fixed gap of 10 mm. Experimental data are compared with the simulation with Ar and 200 ppm $H_2O$ in the plasma gas mixture.***

Figure 11(a) presents the $\tau_{eff}$ values computed from Eq. (6) while varying the air content in the $10^{-3}$% - 100% range, for several $H_2O$ concentrations in the plasma gas mixture. It is noticeable that the latter parameter has a strong influence on the resulting effective lifetime of the LIF signal for air contents lower than 1%.

The evolution of experimental $\tau_{eff}$ as a function of radial position is depicted in Figure 11(b) for several axial positions, for a plasma torch-to-substrate separation distance of 5 mm. It should be noted that the standard error on the exponential decay fit was always smaller than 1.5 ns in the higher range of $\tau_{eff}$ values (and decreased down to 0.2 ns for the shorter $\tau_{eff}$ values presented in Figure 11(d)) and was thus comprised in the limits of the square symbols. The value of $\tau_{eff}$ fluctuates between 155 ns and 215 ns, and the evolution versus radial position is similar for the whole set of axial positions. This fluctuation is not clearly understood. It may arise from some scattering and reflection of the laser light on the exit grid of the plasma torch, due to the short gap. When comparing these experimental data to the simulation, we observe values corresponding to 0% air and spread between 200 and 350 ppm $H_2O$ in the plasma gas mixture. The latter value is similar to the result of computation with 0.2% air and a plasma gas mixture of Ar with 200 ppm $H_2O$. Hence, for a torch-to-substrate separation distance of 5mm, the air content is limited to 0.2% whatever the radial position.

Although we have estimated, through mass measurements of the bubbler, that the $H_2O$ concentration injected in the plasma torch is 370 ppm on average, further determinations of the air content are achieved by comparing the experimental data to the simulation with Ar and 200 ppm $H_2O$ in the plasma gas mixture. Most LIF measurements were carried out at a radial position of 0mm, and these parameters give the best fit for a gap of 5mm, assuming a negligible air content at this location (cf. results of mass spectrometry in Sec. III C).

Figure 11(c) presents the evolution of experimental $\tau_{eff}$ as a function of the plasma torch-to-substrate separation distance at a radial position of 0mm and an axial position of 1.4 mm. The effective lifetime decreases as the substrate is moved away from the plasma torch. A faster decrease is observed for torch-to-substrate separation distances longer than 8 mm, the estimated air content rising from 0.1% for a gap of 8mm to 3%–4% for a gap of 12 mm. Qualitatively, this trend is well correlated to mass spectrometry measurements.

When the torch-to-substrate separation distance is set to 10mm (cf. Figure 11(d)), the effective lifetime value falls as the radial position moves from 0 to 12.5 mm. For an axial position of 1.4 mm, the value of $\tau_{eff}$ stabilizes, which again may arise from interfering with scattered laser light signal. For other axial positions, the shortest $\tau_{eff}$ value that can be measured is 5–10 ns. At a radial position of 12.5mm, the LIF signal intensity is generally too weak and $\tau_{eff}$ cannot be extracted, indicating that the air content exceeds 20%.

# 4. Discussion

## 4.1. Air content between the plasma torch and the substrate

Both mass spectrometry and laser induced fluorescence on OH radicals have been used for spatially resolved measurements of the air content between the plasma torch and the substrate, while varying the torch-to-substrate separation distance. Comparison of the results obtained with the two diagnostics reveals some discrepancies in the values obtained: e.g., for a gap of 10 mm, the air content estimated by LIF is at most 1.4% and 10% at radial positions of 0mm and 6mm, respectively, while mass spectrometry measurements give $N_2$ relative concentrations reaching 5% and 24%, corresponding to air contents of 6.4% and 31% for the same radial positions, respectively. However, the relative trends are similar: the air concentration is negligible (compared to lower limit measurement threshold of mass spectrometry) for gaps shorter than 5 mm, and strongly increases from gaps of 8mm onwards. Radial gradients in the concentrations of species then appear, which is coherent with an effect of air entrainment by the gas flowing out of the plasma torch.

It should be recalled that mass spectrometry is an intrusive diagnostic, and that only the species which are incident on the capillary entrance can be detected. The relative intensities measured are thus critically dependent on the positioning of the capillary. Moreover, the detector is more sensitive to ions of lower m/z ratio, which could participate to the higher $N_2$ concentrations measured by this method. Finally, the relative concentrations have been calculated by considering only Ar, N2, $O_2$, $H_2O$, and $CO_2$.

Laser induced fluorescence is practically a non intrusive technique in the linear excitation regime, where the level depopulated by laser pumping is instantaneously refilled through RET. It allows spatially resolved measurements with the implementation of an appropriate optical set-up (radial resolution of 1mm and axial resolution of 0.8mm were achieved in this work). In our experimental conditions, the air concentration values obtained from the comparison of experimental and simulated effective lifetime of $OH(A^2\Sigma^+; v'=0)$ radicals should be more reliable than the values obtained from mass spectrometry measurements when the air content is in 1%–15% range. Beyond a concentration of 15%, the effective lifetime is shorter than 7 ns and the recorded LIF signal intensity is too weak. The two techniques employed are thus complementary for the determination of the air content over a wide range.

## 4.2. Influence of ambient air on the flowing afterglow

The air entrainment under the plasma torch strongly modifies the properties of the flowing afterglow of the $Ar/O_2$ plasma. In the present study, this results in emission from $N_2(C)$ molecules being observed, the Ar excitation temperature $T_{13}$ going down, and ozone molecules being created, provided $O_2$ is injected in the plasma torch.

The emergence of $N_2$ (C–B) emission is correlated to the sharp increase in air concentration as measured by LIF and mass spectrometry. Possible pathways for the excitation of $N_2$ molecules to the $C^3\Pi_u$ electronic state include: electron impact excitation from lower excited states or from the $X^1\Sigma_g^+$ ground state, energy pooling reactions between $N_2(A^3\Sigma_u^+)$ molecules,38 cascade transitions from higher excited states, and collision induced energy transfer between Ar metastable atoms and $N2(X^1\Sigma_g^+)$ molecules.39 The electron density and Ar metastable concentration determined from a two-dimensional fluid model on a very similar plasma source (the excitation frequency is 13.56 MHz, a pure Ar discharge is considered, and there is no inlet distributor under the grounded electrode) lie in the range $10^{10}$–$10^{11}$ $cm^{-3}$ and $10^9$–$10^{10}$ $cm^{-3}$, respectively.40 Stepwise electron impact excitation, energy pooling reactions in N2(A), and energy transfers between Ar metastable atoms and $N_2(X)$ molecules may contribute significantly to the population of $N_2(C)$ state.

The rise in $N_2$ (C–B) emission intensity with increasing plasma torch-to-substrate separation distance up to 9mm is reasonably expected from the rise in $N_2$ concentration. The subsequent decrease in emission intensity for higher separation distances can be assigned to the increasing collisional quenching rate of $N_2(C^3\Pi_u; v'=0)$ state by $N_2$, $O_2$, and $H_2O$ molecules from the ambient air. The average rate constants for quenching of $N_2(C^3\Pi_u; v'=0)$ state by $N_2$, $O_2$, $H_2O$, and Ar found in the literature are $1.14*10^{-11}$ $cm^3s^{-1}$,41 $3.0*10^{-10}$ $cm^3.s^{-1}$,42 $3.9*10^{-10}$ $cm^3.s^{-1}$,42 and at most $3.3*10^{-12}$ $cm^3$ $s^{-1}$,43 respectively. The average radiative frequency is high ($2.77*10^7$ $s^{-1}$ (Ref. 41)), but at atmospheric pressure and at a gas temperature of 400 K, we estimate that the collisional quenching rate is already twice as high as the radiative frequency in an argon environment. Using the same air composition as in Sec. IIID, and a mixture of Ar with 200 ppm of H2O for the plasma gas composition, the estimated total quenching rate varies from 6.1*$10^7$ $s^{-1}$ to 1.3*$10^8$ $s^{-1}$ for a radial position of 0mm, as the torch-to-substrate separation distance increases from 5mm to 12 mm. At a radial position of 12.5 mm, the total quenching rate exceeds 1.3*$10^9$ $s^{-1}$ for gaps longer than 10 mm. It is therefore very likely that the progressive decay in the (averaged over the line of sight) concentration of $N_2$(C) molecules for increasing torch-to-substrate separation distance above 9mm is related to their increasing quenching rate by $N_2/O_2/H_2O$.

The continuous decrease in Ar excitation temperature $T_{13}$ (close to the exit grid of the plasma torch), hence in Ar 4 p population, with increasing torch-to-substrate separation distance, can be explained by a combination of diminishing Ar concentration due to air entrainment, and growing quenching rate of Ar 4 p states by $N_2$ and $O_2$ molecules (in the absence of data for collisional quenching by $H_2O$ molecules).

The rate constants for quenching of Ar 4 p states by $N_2$, $O_2$, and Ar are, depending on the 4 p state considered, 0.3–3.5*$10^{-10}$ $cm^3.s^{-1}$, 4.6–7.6*$10^{-10}$ $cm^3.s^{-1}$, and 1.2–1.6*$10^{-11}$ $cm^3.s^{-1}$, respectively.44 In pure Ar plasma, at atmospheric pressure and T=400 K, the quenching rate of Ar 4 p states by Ar atoms is 2.1-2.9*$10^8$ $s^{-1}$, i.e., about ten times larger than the radiative frequency (2.5–4.5*$10^7$ $s^{-1}$ (Ref. 45)). The quenching mechanisms of the Ar 4p states by collision with Ar atoms are relaxation to other 4p states or lower 4s states, whereas the quenching mechanisms for collisions with $N_2$ and $O_2$ are excitation-transfer reactions,44 with Ar atoms decaying back to the electronic ground state. The total quenching rate of Ar 4p states changes from 2.2–2.9*$10^8$ $s^{-1}$ to 2.2-4.5*$10^8$ $s^{-1}$ for a radial position of 0mm, as the torch-to-substrate separation distance increases from 5mm to 12mm. At a radial position of 12.5 mm, the total quenching rate is above 0.4–4.6*$10^9$ $s^{-1}$ for gaps bigger than 10 mm.

Moreover, Ar metastable states are also quenched through excitation-transfer reactions with $N_2$(X) molecules, leading to emission in the $N_2$(C–B) system. Since the Ar 4 p states are in great part populated by electron collisions with Ar 4s atoms,46 an increase in the quenching rate of metastable states by $N_2$ will in turn reduce the rate of formation of Ar 4 p states and participate in the observed decay in their concentration.

Absorption measurements have revealed that, on average, ozone forms at concentrations larger than $10^{14}$ $cm^{-3}$ when both the torch-to-substrate separation distance exceeds 8mm and $O_2$ is added to the inlet Ar gas flow. Since the ozone formation rate is controlled by three-body recombination involving $O_2$ and O species, this can signify that the dissociation rate of $O_2$ in the plasma discharge and afterglow is so high that extra $O_2$ molecules are required to reach a detectable ozone concentration; these are brought by the entrained air. Another possibility is that entrained $N_2$ molecules, excited in the afterglow to electronic states $A^3\Sigma_u^+$ and $B^3\Pi_g^+$, transfer their energy through collisions to $O_2$ molecules, to give rise to O atoms which then participate in reactions for ozone generation.47 Measurements of the absolute concentration of O radicals must be undertaken to gain insight into the reaction kinetics of ozone creation.

## 4.3. Implications for surface treatment

The AtomfloTM 250D plasma source was designed to operate in the open air for polymer surface treatment and PECVD. Our experimental results indicate that the air content in between the plasma torch and the substrate is negligible for torch-to-substrate separation distances shorter than 5 mm. NH(A) molecules have been detected, but the absence of emission in the $N_2$(C–B) system suggests that $N_2$ molecules are completely dissociated, which further supports the conclusion that the air diffusion is very weak. Therefore, for small torch-to-substrate separation distances, the flowing afterglow creates a protective atmosphere above the substrate, and the species incident at the surface are controlled by the plasma operating conditions only. When heat sensitive materials such as polyethylene need to be treated, separation distances greater than 5mm are often chosen,6 the gas temperature being approximately 400K close to the exit grid. In that case, the influence of ambient air can no longer be neglected, and radial gradients in the concentrations of neutral species appear, which may lead to spatial inhomogeneities for surface treatments. Furthermore, new species are generated from the interaction of the flowing afterglow and the entrained air: ozone is detected at average concentrations higher than $10^{14}$ $cm^{-3}$ when $O_2$ is injected with Ar. Apart from O and OH radicals, which are formed in the plasma discharge region and the afterglow, it is highly probable that other reactive oxidative species are also created, such as NOx species. These are susceptible to play a role in the surface modification of polymers.

# 4. Conclusion

The influence of ambient air on the flowing afterglow of an atmospheric pressure $Ar/O_2$ plasma (with water vapor injected downstream of the electrodes) has been investigated by means of optical diagnostics and mass spectrometry. Both laser induced fluorescence on OH radicals and mass spectrometry were implemented to determine the air content in between the plasma torch and the substrate, while varying the torch-to-substrate separation distance from 2mm to 12 mm. Our measurements indicate that for short separation distances (<5 mm), no air is detected (within the limits of measurement uncertainties), hence the nature of species incident at the substrate surface is controlled by the plasma operating conditions. For separation distances larger than 8mm, the air content dramatically increases, and radial gradients in the concentrations of species are measured, consistent with an effect of air entrainment by the gas flowing out of the plasma torch. The air entrainment induces a diminution in the Ar 4 p population, along with the appearance of optical emission in the $N_2$(C–B) system. The decrease in Ar 4 p and $N_2$(C) concentrations for separation distances larger than 9mm are assigned, at least in part, to the increasing rate of collisional quenching by $N_2$ and $O_2$ molecules from the ambient air. $O_3$ molecules are also detected at concentrations higher than $10^{14}$ $cm^{-3}$ for torch-to-substrate distances higher than 8 mm, provided $O_2$ is injected with Ar. This result leads us to expect that other reactive oxidative species, such as NOx, are generated from the interaction of the flowing $Ar/O_2$ plasma afterglow with the ambient air.

# 5. Acknowledgements

This work was carried out in the framework of the Interuniversitary Attraction Pole program “Plasma Surface Interactions” financially supported by the Belgian Federal Office for Science Policy (BELSPO). The authors gratefully acknowledge the people from the Department of Applied Physics-Research unit Plasma Technology, Ghent University, Belgium, for lending their calibrated tungstenhalogen lamp and for useful advice. The people from the “Service de Chimie Quantique et Photophysique,” Faculty of Sciences, Université Libre de Bruxelles, Belgium are also acknowledged for lending their mercury lamp.